\documentclass[12pt]{article}
\usepackage{times}
\usepackage{geometry}
\usepackage[utf8]{inputenc}
\usepackage{enumitem,amssymb}
\usepackage{ragged2e}
\newlist{thematic}{itemize}{8}
\setlist[thematic]{label=$\square$}
\usepackage{pifont}

\usepackage{graphics}
\usepackage{epsfig}
\usepackage{wrapfig}

\begin{document}
\raggedright

\huge
Astro2020 Science White Paper \linebreak

Do Supermassive Black Hole Winds Impact Galaxy Evolution? \linebreak
\normalsize

\noindent \textbf{Thematic Areas:} Formation and Evolution of Compact Objects; Galaxy Evolution; Multi-Messenger Astronomy and Astrophysics
 
\vspace{0.3cm}
 
\textbf{Principal Author:}

Name: Dr. Francesco Tombesi	
 \linebreak						
Institution: University of Maryland, College Park; NASA/GSFC;
University of Rome ``Tor Vergata'' 
 \linebreak
Email: ftombesi@astro.umd.edu
 \linebreak
 
\textbf{Co-authors:} M. Cappi (INAF-OAS, Italy), F. Carrera (Instituto de Fisica de
  Cantabria, Spain), G. Chartas (College of Charleston, USA),
  K. Fukumura (James Medison University, USA), M. Guainazzi (European
  Space Agency, Netherlands), D. Kazanas (NASA/GSFC, USA), G. Kriss (Space Telescope Science
  Institute, USA), D. Proga (University of Nevada, Las Vegas, USA),
  T. Turner (University of Maryland, Baltimore County, USA),
  Y. Ueda (Kyoto University, Japan), S. Veilleux (University of
  Maryland, College Park), M. Brusa (University of Bologna, Italy),
  M. Gaspari (Princeton University, USA)
  \linebreak

   \begin{figure}[h]
   \centering
   \includegraphics[width=14cm,height=8.5cm,angle=0]{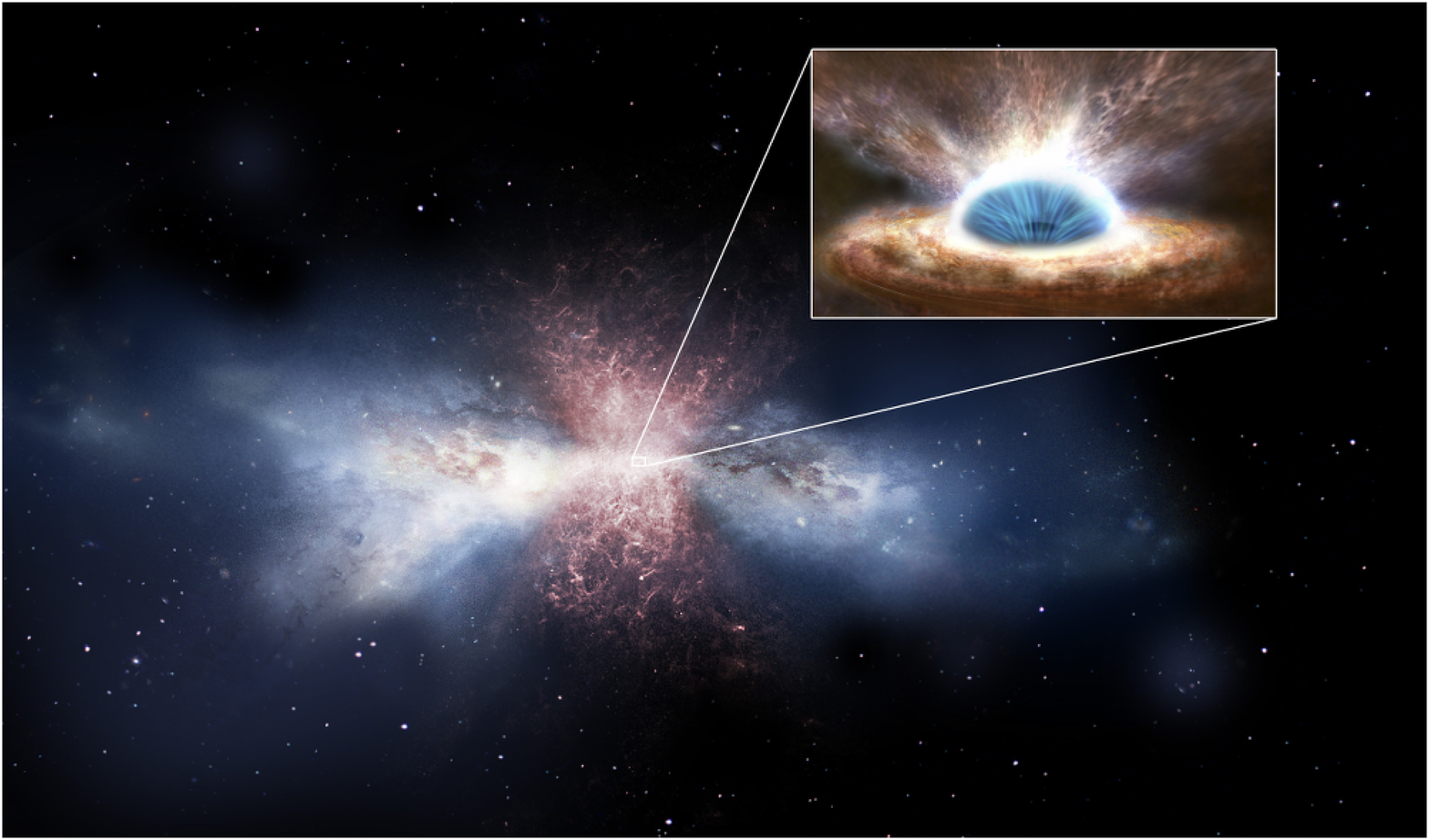}
     \end{figure}

\pagebreak

\justify

{\noindent \bf {\underline{Introduction}}}

Virtually every galaxy in the universe hosts a supermassive black hole
(SMBH) at its center, with masses that are found to correlate with
galactic-scale properties (e.g., Kormendy 2016). The growth of SMBHs may be triggered by both galaxy mergers in the
early universe and/or by secular processes internal to their host
galaxies, mostly observed in low-redshift isolated sources (e.g.,
Kormendy \& Kennicutt 2004). As dust and gas are gradually dispersed from the central
regions,  a  completely obscured active galactic nucleus (AGN) evolves to a dusty phase and finally to a
completely exposed quasar (e.g., Hopkins et al.~2006). Galactic winds driven by the central AGN have been invoked to play a fundamental role in
this phase, quenching the growth of both the SMBH and stellar
spheroidal component, and explaining the tight SMBH-spheroid mass relation (Silk \& Rees 1998;
King \& Pounds 2003). Moreover, AGN
outflows, in the form of winds and jets, are also required to prevent
catastrophic cooling flows and
for dispersing metals at galactic and intergalactic distances (e.g.,
Gaspari \& Sadowski 2017a). Understanding the physics of how mass and energy flows over such
  enormous spatial scales is of paramount importance to understand
  both galaxy and SMBH evolution. 

Within this context, high-velocity winds in the X-ray band are regarded as
the most effective way of transporting energy from the nuclear 
to galactic scales (Tombesi et al. 2010, 2015; Nardini et
al.~2015; King \& Pounds 2015). According to models, the energy of
such SMBH driven winds is deposited into the host galaxy
interstellar medium (ISM), resulting into the recently observed
galactic-scale molecular outflows, which are able to sweep away the
galaxy's reservoir of gas and quench star formation activity
(e.g., Zubovas \& King 2012; Faucher-Giguere \& Quataert 2012). 
\underline{The key questions to be addressed here are:} {\it How do SMBHs launch winds/outflows,
  what is their physical geometry, and do they affect SMBH growth?}
{\it How are the energy and metals transferred and deposited into the
  galactic and circumgalactic medium?} {\it What is the relation between
  SMBH feeding and feedback?} {\it How does this self-regulated cycle develops?}

X-ray observations are a unique way to address these questions because
they probe the initial, highest-ionization and hottest phase of the outflow, which carries most of the kinetic
energy (e.g., Tombesi et al.~2013). High throughput, high spectral and spatial resolution
instruments as those proposed for
\emph{Athena}\footnote{https://www.the-athena-x-ray-observatory.eu/}
(Barcons et al.~2017),
\emph{Lynx}\footnote{https://wwwastro.msfc.nasa.gov/lynx/} ({\"O}zel
2018), and \emph{AXIS}\footnote{http://axis.astro.umd.edu/} (Mushotzky 2018) are required to
determine the acceleration and launching mechanism(s) of SMBH driven outflows and to quantify their
impact on galaxy evolution.\\

{\noindent \bf {\underline{How do SMBHs drive winds and how much mass
      and power do they carry?}}}

SMBH winds are directly observed as blue-shifted and
broadened absorption lines in the X-ray spectra of at least half of
AGNs (e.g., Crenshaw \& Kraemer 2012; Tombesi et al. 2013). These
absorption systems span a range of about six orders of magnitude of velocities and physical
conditions, suggesting a continuous stream that carries kinetic energy and
momentum away from sub-pc scales potentially up to the kpc-scale
environment (e.g., Arav et al.~2018). Simulations of accretion disks and outflows have
progressed enormously in recent decades, and show that several
physical mechanisms (thermal, radiation, magnetic)
are able to accelerate winds, providing a
theoretical basis for understanding the observations (e.g., Blandford \& Payne 1982; King \&
Pounds 2003; Proga, Stone \& Kallman 2000; Fukumura et
al.~2010). However, what determines the dominant mechanism is not yet understood.

%These warm absorber carry only a small fraction of the
%kinetic power, as the amount of material and outflow velocity are both
%relatively small. Instead, there are two much higher velocity types of
%outflows which potentially have much greater impact on the host
%galaxy.

%They may probably arise in a UV line
%driven wind from the accretion disk (e.g., Proga \& Kallman 2004).

   \begin{figure}[t]
   \centering
  \includegraphics[width=8cm,height=6.5cm,angle=0]{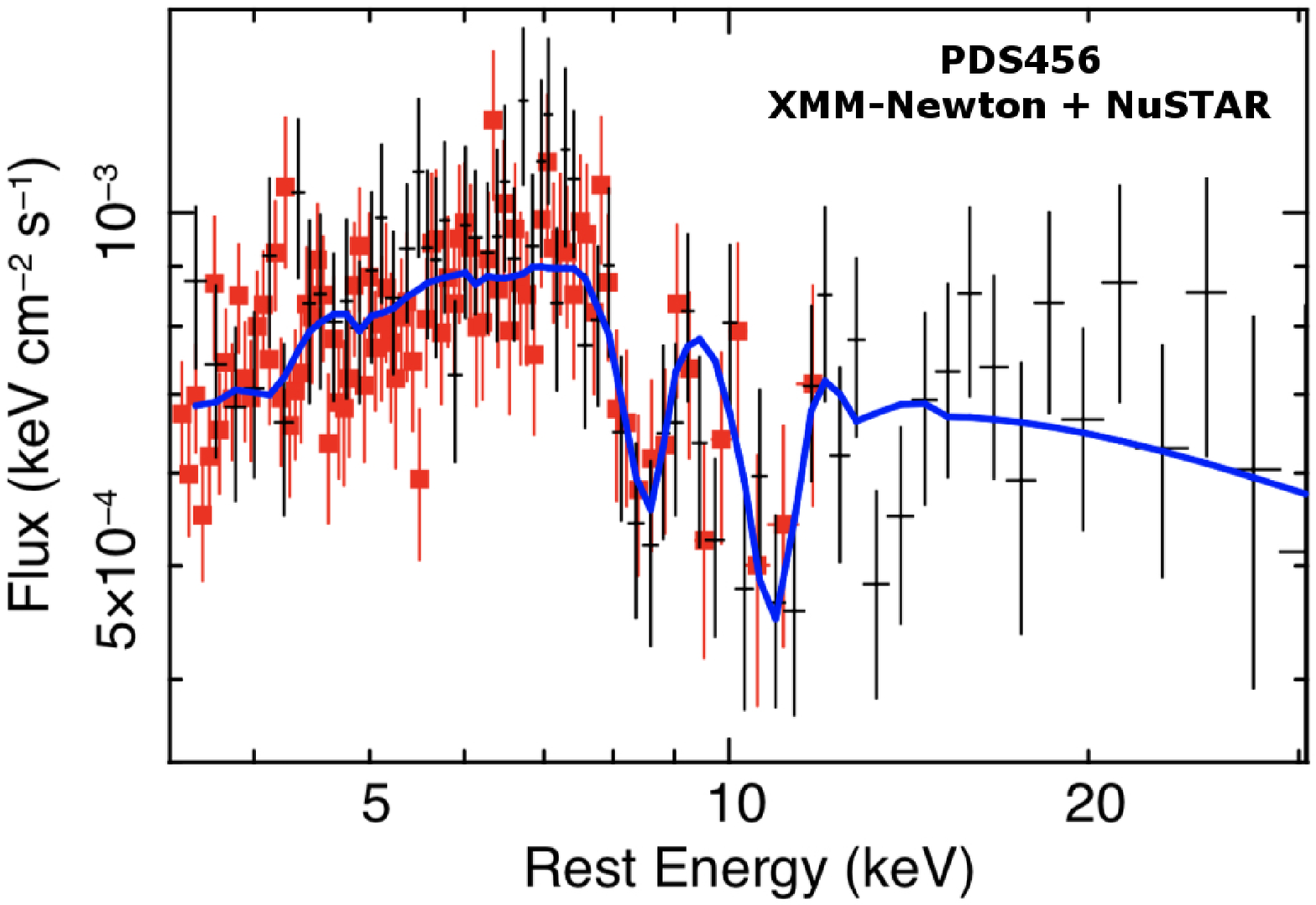}
\hspace{0.1 cm}
   \includegraphics[width=8cm,height=7cm,angle=0]{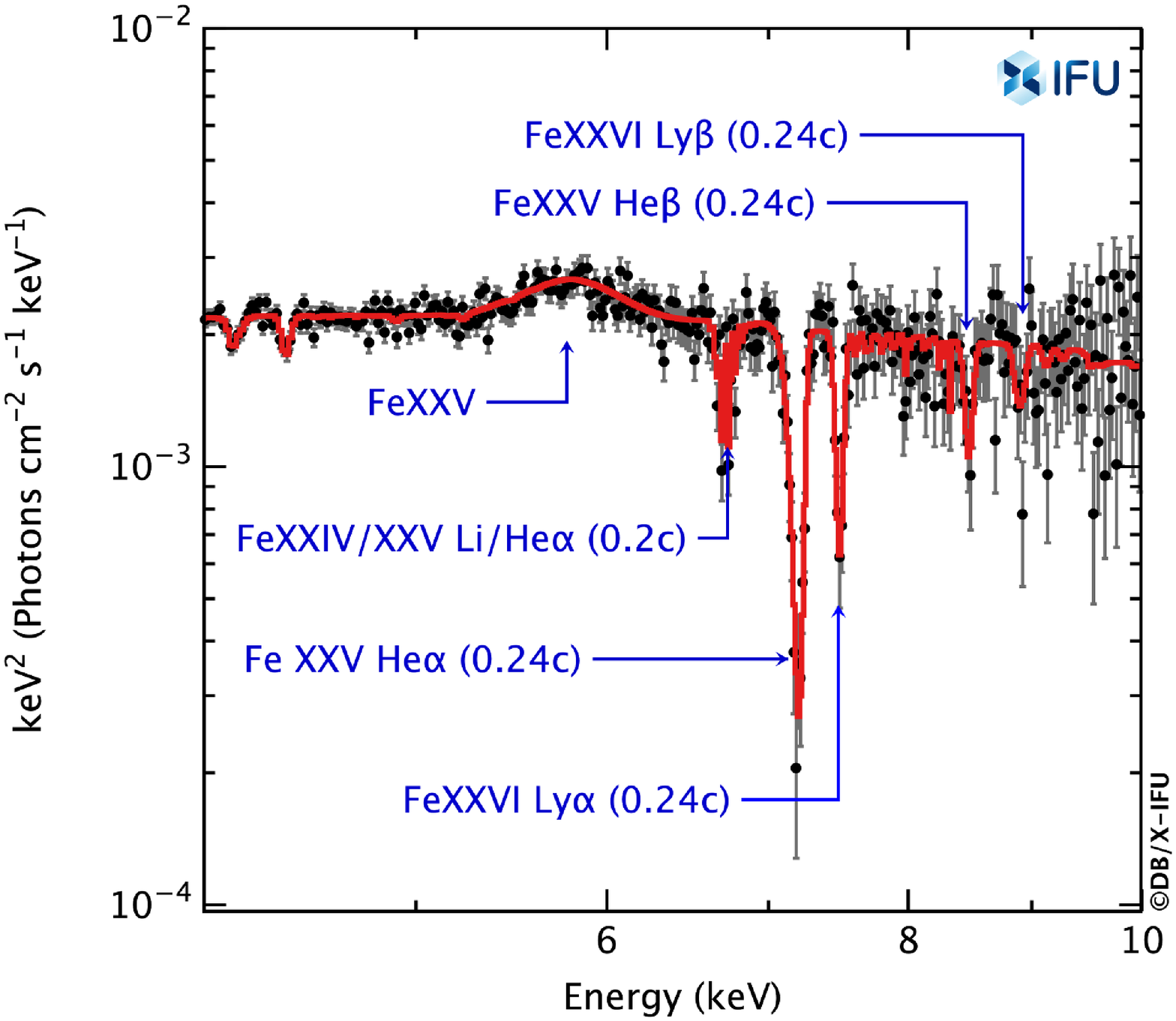}
   \caption{\footnotesize{{\it Left:} Simultaneous 150~ks XMM-Newton
       \& NuSTAR spectra of the quasar PDS~456 showing hints of a
       highly ionized outflow with two relativistic velocity
       components in absorption (Reeves et al.~2018). {\it Right:} Simulated 100 ks \emph{Athena} X-IFU
       spectrum of the same source. A series of absorption lines from
       an outflow with two velocity components at $v_{out}$$=$0.20--0.24c and a turbulent velocity broadening of
       3,000 km~s$^{-1}$ would be clearly detectable thanks to the
       unprecedented high-energy resolution
       and throughput provided by the \emph{Athena} X-IFU (Credits: X-IFU Consortium).}}
     \end{figure}

Blue-shifted narrow absorption lines in the UV and soft X-rays suggest
outflows with moderate velocities of hundreds to few thousands
km/s. These ``warm absorbers'' are detected in $>$50\% of AGN
(Crenshaw \& Kraemer 2012), and may have an origin in the swept-up
interstellar medium (ISM) or thermally driven winds from the outer
accretion disk. 
In the UV band, broad absorption lines are seen in $\sim$30\%
of AGN, and may be present outside the line of sight in most quasars
(Ganguly \& Brotherton 2008). These absorbers can be outflowing with
velocities as high as $\sim$20\% of the speed of light, and so they carry
considerable kinetic power, defined as
$E_k=(1/2)\dot{M}_{out}v_{out}^2$, where $\dot{M}$ is the mass outflow
rate. 

The most powerful observed outflows appear to be so highly ionized
that only the bound transitions of hydrogen- and helium-like
iron are left, making them detectable only at hard X-ray energies. These
X-ray winds are observed in
$>$30\% of local AGN, and even in some higher
redshift quasars (Chartas et al. 2002; Lanzuisi et al. 2012), with outflow velocities of up to $\sim$30\% of
the speed of light (Tombesi et al. 2010). These
``ultra-fast outflows'' (or UFOs) have velocities that point to an
origin very close to the SMBH, but the launching and acceleration
mechanism(s) remain unclear. 
%Possibilities include radiation-driven winds and/or magnetic
%driving, yet we are still searching for the observational answers to
%this basic question: how do accreting SMBHs launch winds/outflows?

The key to progress on this investigation is a detailed characterization of
the physical properties of these winds (column density, ionization
state, outflow velocity, location, geometry, covering factor,
etc.). The upcoming X-ray Imaging and Spectroscopy Mission
(\emph{XRISM}) will provide high spectral resolution observations but,
due to the relatively low collecting area and spatial resolution,
these will be limited to the nearby brightest AGNs (e.g., Kaastra et al.~2014).
Only the high-energy resolution and high throughput offered by
the proposed \emph{Athena} (e.g., Cappi et al.~2013) and \emph{Lynx}
(e.g., {\"O}zel 2018) X-ray observatories will
allow the study of such outflows on a large enough sample of sources
to effectively probe the prevalence and evolution of these systems from the early
universe to the current epoch (e.g., Georgakakis et al.~2013). These
proposed missions will provide enough S/N to utilize short time-scale
($\simeq$ hrs) variability as a tool to explore the wind launching
region down to a few Schwarzschild radii (e.g., Miller
et al.~2010; Waters et al.~2017).  

The unprecedented data quality will allow us to seek correlations among the fundamental parameters such as
density, ionization, outflow velocity, and luminosity, which
will uniquely constrain predictions of radiation-driven (e.g., Proga, Stone \& Kallman 2000), momentum-driven (King 2010), and
magnetically-driven (Fukumura et al. 2010) accretion disk wind
models. With such next generation data we will also be able to quantify the outflow
efficiency (i.e., the ratio between the accretion and ejection rates),
as well as the kinetic power budget of the various components in the
wind to better assess their impact on the large-scale environment of
the host galaxy. The feasibility of such analysis is shown in Fig.~1,
where the spectrum of the quasar PDS~456
obtained simultaneously with \emph{XMM-Newton} and \emph{NuSTAR}
showing hints of a possible two velocity component outflow (Reeves et
al.~2018) is compared to simulated data from the
\emph{Athena} X-ray integral field unit (X-IFU), which will allow a
full determination of the outflow kinematic and ionization structure,
necessary for reliably constraining the wind energetics. 

Such X-ray observations are key for determining the total
column density, the highest velocities, the highest ionization, and
hence the kinetic power carried by SMBH outflows, which is required to drive the ensuing galaxy-scale outflows. The proposed \emph{Athena}
and \emph{Lynx} high spectroscopic throughput will allow for a giant leap in sensitivity to
most ionization states of light elements, and to all those of iron
(from neutral to hydrogen-like), hence providing a detailed characterization and
understanding of the SMBH outflow.\\

{\noindent \bf {\underline{How do SMBH winds interact with the
    galaxy-scale environment?}}}

Understanding accretion and ejection processes is fundamental to uncover the
origin of AGN feedback and the co-evolution of SMBHs and their host
galaxies. Depending on the covering factors and duty cycles, the mass
outflow rate from SMBH outflows may reach kinetic powers of a few per
cent of the AGN bolometric luminosity, exerting a significant impact
on the host galaxy, as shown by numerical simulations (Hopkins \& Elvis
2010; Ostriker et al. 2010; Gaspari, Brighenti \& Temi 2012; Zubovas \& King 2012; Faucher-Giguere \& Quataert 2012; Wagner et al. 2013; Gaspari \& Sadowski 2017a). See the
left panel of Fig. 2 for a schematic view. 

%Over time, AGN outflows are likely to significantly impact the galaxy
%bulge evolution, star formation, and SMBH growth consistent with the
%observed SMBH-host galaxy relationships (e.g., Silk \& Rees 1998;
%Fabian 1999; Ferrarese \& Merritt 2000; King 2010; Ostriker et
%al. 2010; Zubovas \& King 2012; Faucher-Giguere \& Quataert 2012; Gaspari \& Sadowski 2017). Other cosmologically important issues in which AGN winds
%and feedback are likely to play an important role are in: i) limiting
%the upper mass of galaxies by halting galaxy growth; ii) providing
%the extra heating in galaxy cluster cooling flows, especially when AGN activity is also
%related to relativistic jets; iii) contributing with “super-winds”
%from starburst galaxies to the enrichment of the intergalactic medium,
%and iv) preventing star formation in mergers by effectively removing
%the coldest ISM gas from galaxies.

While fast AGN outflows can carry a large amount of mechanical power
from the innermost region nearest to SMBHs, how, if, and where this
energy is actually released into the interstellar medium is far from
understood. Although there is wide consensus that AGN feedback is
likely able to quench star formation, a quantitative understanding of
how this process works is still lacking. A breakthrough will come only
from coordinated synergies between 
high-quality X-ray data (probing the first phase of the wind) and lFU observations at longer wavelengths (probing the mass and extension of the molecular and ionised outflows).

   \begin{figure}[t]
   \centering
   \includegraphics[width=8cm,height=7cm,angle=0]{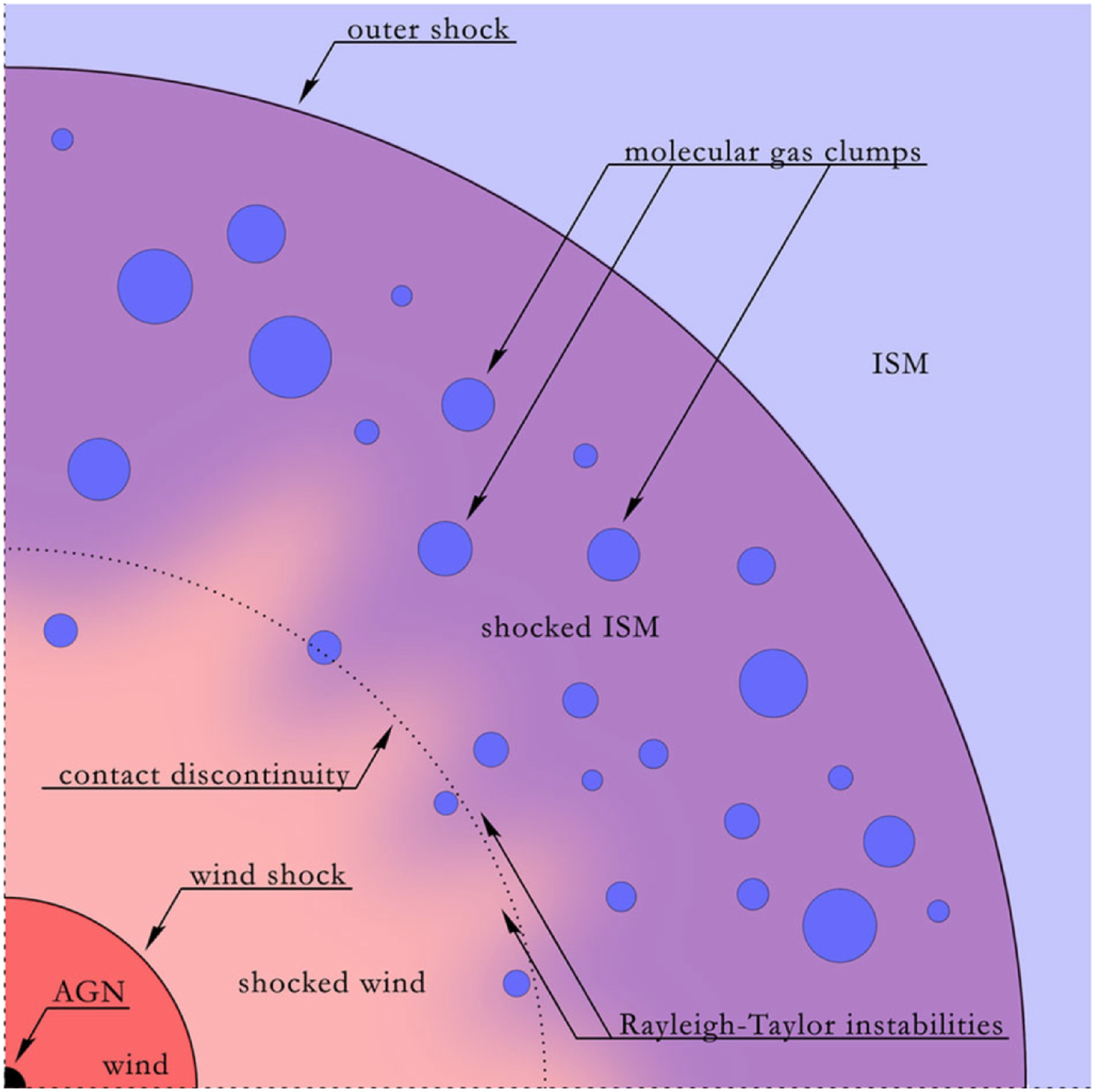}
\hspace{0.1 cm}
   \includegraphics[width=8cm,height=7cm,angle=0]{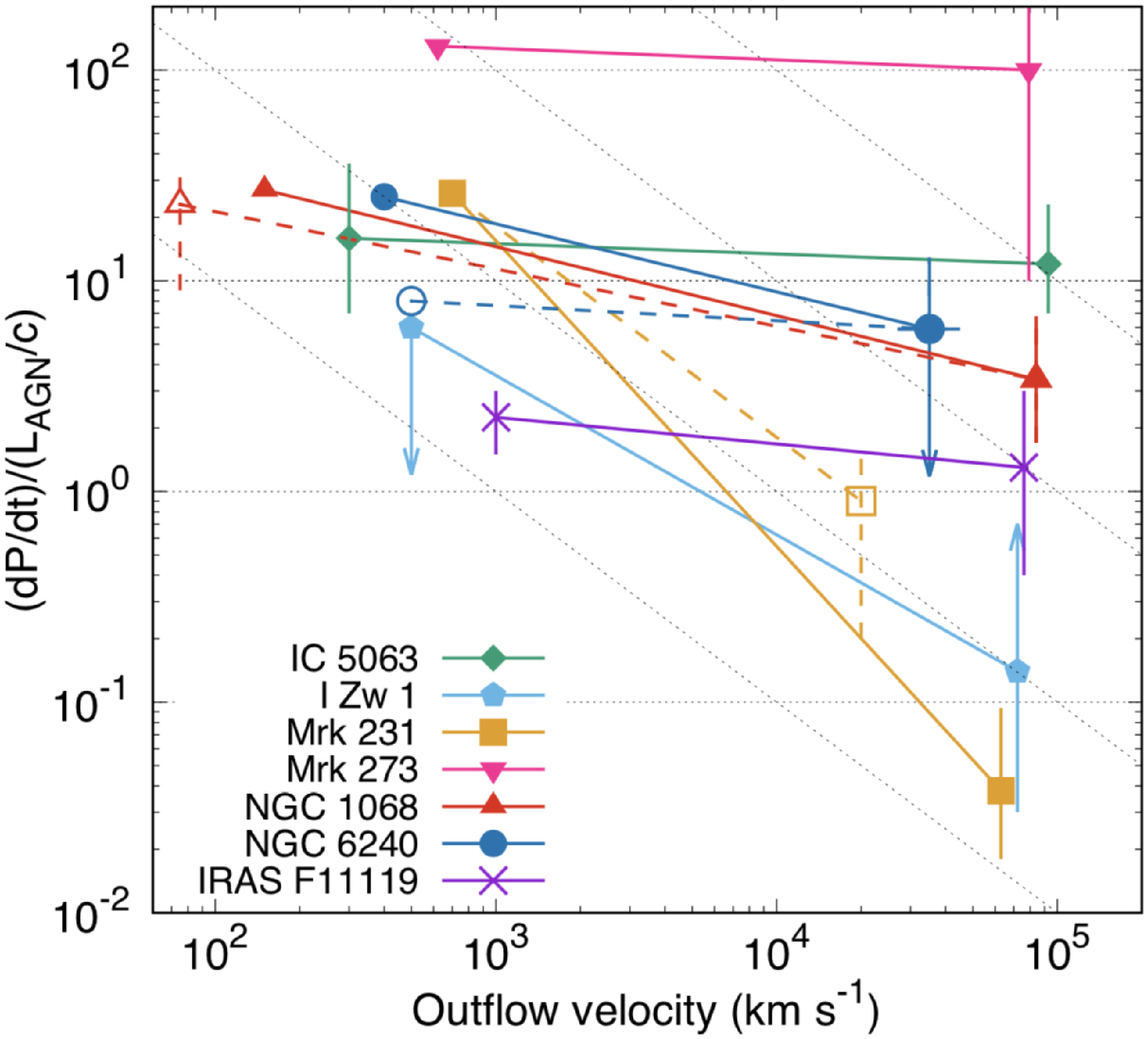}
   \caption{\footnotesize{\emph{Left:} Schematic view of
       the multi-phase AGN driven outflow launched near the central SMBH
       and interacting with the host galaxy interstellar medium
       (Zubovas \& King 2014). \emph{Right:} Momentum flux versus
       outflow velocity for the inner SMBH outflows (right side) and
       galaxy-scale molecular outflows (left side) in a small sample
       of seven active galaxies. The horizontal
       lines show the momentum-driven flows, whereas the ones
       ascending toward the left show the energy-driven flows
       (Mizumoto et al.~2018).}}
     \end{figure}

Observational evidence is mounting that AGN act on the multi-phase -
molecular, atomic, ionized - ISM of their host galaxies on large
scales and, through this,
affect the evolution and transformation of the host
galaxy (e.g., Cicone et al.~2018). For instance, kpc-scale outflows
have recently been discovered in both star-forming galaxies and in powerful AGN
(Veilleux et al. 2013), at
millimeter and far-infrared wavelengths (with IRAM, ALMA,
Herschel). The inferred mass outflow rates of up to several hundreds
of solar masses per year and high kinetic power suggest that AGN
activity is the driver, providing strong support
 to models of AGN feedback. Evidence of powerful, massive molecular
 outflows has also been found in a few quasars at high redshifts (Maiolino
 et al. 2012).

The first evidence of a SMBH wind with a
velocity of 25\% of the speed of light in a late-stage galaxy merger hosting a massive
molecular outflow was found comparing the X-ray and F-IR spectra of IRAS
F11119$+$3257  (Tombesi et al. 2015). The comparison of the energetics of
these outflows indicates a connection between the SMBH
driven wind and the large-scale molecular outflow. This is shown for an
additional six sources in the right panel of Fig.~2 from Mizumoto et al.~(2018).
Such a diagram provides a way to trace the SMBH
wind observed in X-rays to the depletion of the “fuel” required for
star formation activity, outlined by the large-scale molecular outflow
observed in infrared and mm spectra (Tombesi et al. 2015; Feruglio et
al.~2015, 2017; Veilleux et
al.~2017). Clearly, there is an urgent need to
populate the plot in the right side of Fig.~2 with a larger number of higher quality data points in order to directly compare it with theoretical models and estimate the global parameters of this crucial astrophysical process. 

In nearby AGN, feedback can be directly probed by 3-D mapping of the various components which overlap spectrally or
spatially. Along with spectroscopy, good spatial resolution is therefore  necessary
for a full characterization of the winds phenomenon. In this respect the few arcsec resolution of
\emph{Athena} is the minimum needed to start
resolving the outflow patterns traced by filaments, bubbles, shells,
streamers, etc. in nearby active galaxies (Wang et al.~2011; Paggi et al.~2017). This can be pushed to much
higher detail and higher redshifts with the sub-arcsec resolution
envisioned for \emph{AXIS} and \emph{Lynx}, close to the resolution
achievable with \emph{JWST} and \emph{ALMA}.

   \begin{figure}[t]
   \centering
   \includegraphics[width=16cm,height=9cm,angle=0]{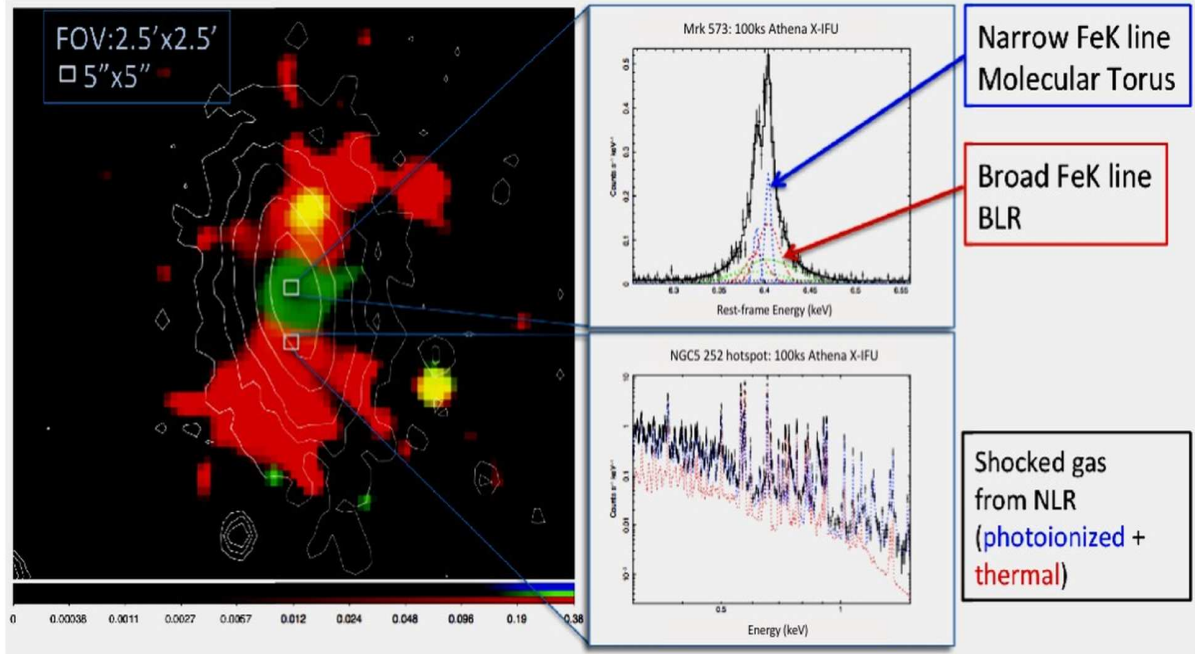}
   \caption{\footnotesize{\emph{Left:} \emph{Athena} simulated color-coded image of the nearby Seyfert 2 galaxy NGC 5252 (from the Chandra image in Dadina et al. 2010). The soft X-rays are known to trace the [OIII] ionization cones forming a biconical outflow/illumination pattern driven by the AGN which impacts all over the host galaxy (DSS optical contours indicated in white). \emph{Right-top:} \emph{Athena} X-IFU simulated spectrum of the Fe~K emission line assuming the sum of a broad plus a narrow line component, from the BLR and molecular torus respectively. \emph{Right-bottom:} \emph{Athena} X-IFU simulated spectrum of a part of the plume of ionized emission south of the nucleus and attributed to 25\% of shock (thermal) emission plus 75\% of photoionized emission. Adapted from Cappi et al.~(2013).}}
     \end{figure} 

Studies have been performed to trace the
large-scale soft X-ray emission in dozens of nearby AGN, on scales
ranging from a few arcsec to several tens of arcsec (Wang et
al. 2011, 2012). However, to date, determining whether the extended
soft X-ray emission and AGN-excited narrow line region may be related
to an outflowing wind and/or shocks has been
hampered by insufficient spatial resolution and sensitivity  (Pounds \& Vaughan 2011; Fischer et al. 2013). Fig. 3 illustrates the capabilities of an X-IFU, as proposed on
\emph{Athena}, for studying AGN feedback in nearby Seyfert galaxies.
High spatial and energy resolution are needed to simultaneously resolve the most prominent
components of soft-X-ray diffuse emission, and to spectrally map both
the nuclear and extended outflow components. Finally, it is important to note that feeding and feedback are two
intimately linked phases for SMBH and galaxy co-evolution and
high-resolution, multi-wavelength imaging and spectroscopic
observations, as offered by upcoming IFU detectors, are required to
directly compare with numerical predictions (e.g., Gaspari \& Sadowski
2017a, b; Gaspari et al.~2018).

\pagebreak
\textbf{References}
\bibitem[Arav et al.(2018)]{2018ApJ...857...60A} Arav, N., Liu, G., Xu, X., et al.\ 2018, ApJ, 857, 60 
\bibitem[Hopkins et al.(2006)]{2006ApJS..163....1H} Hopkins, P.~F., Hernquist, L., Cox, T.~J., et al.\ 2006, ApJ, 163, 1 
\bibitem[Silk \& Rees(1998)]{1998A&A...331L...1S} Silk, J., \& Rees, M.~J.\ 1998, A\&A, 331, L1 
\bibitem[King \& Pounds(2003)]{2003MNRAS.345..657K} King, A.~R., \& Pounds, K.~A.\ 2003, MNRAS, 345, 657 
\bibitem[Gaspari \& S{\c a}dowski(2017)]{2017ApJ...837..149G} Gaspari,
  M., \& S{\c a}dowski, A.\ 2017a, ApJ, 837, 149 
\bibitem[Gaspari et al.(2017)]{2017MNRAS.466..677G} Gaspari, M., Temi,
  P., \& Brighenti, F.\ 2017b, MNRAS, 466, 677 
\bibitem[Gaspari et al.(2018)]{2018ApJ...854..167G} Gaspari, M., McDonald, M., Hamer, S.~L., et al.\ 2018, ApJ, 854, 167 
\bibitem[Ganguly \& Brotherton(2008)]{2008ApJ...672..102G} Ganguly, R., \& Brotherton, M.~S.\ 2008, ApJ, 672, 102 
\bibitem[Georgakakis et al.(2013)]{2013arXiv1306.2328G} Georgakakis, A., Carrera, F., Lanzuisi, G., et al.\ 2013, arXiv:1306.2328 
\bibitem[Crenshaw \& Kraemer(2012)]{2012ApJ...753...75C} Crenshaw, D.~M., \& Kraemer, S.~B.\ 2012, ApJ, 753, 75 
\bibitem[Miller et al.(2010)]{2010MNRAS.403..196M} Miller, L., Turner, T.~J., Reeves, J.~N., et al.\ 2010, MNRAS, 403, 196 
\bibitem[Tombesi et al.(2010)]{2010A&A...521A..57T} Tombesi, F., Cappi, M., Reeves, J.~N., et al.\ 2010, A\&A, 521, A57 
\bibitem[Tombesi et al.(2013)]{2013MNRAS.430.1102T} Tombesi, F., Cappi, M., Reeves, J.~N., et al.\ 2013, MNRAS, 430, 1102 
\bibitem[Tombesi et al.(2015)]{2015Natur.519..436T} Tombesi, F., Mel{\'e}ndez, M., Veilleux, S., et al.\ 2015, Nature, 519, 436 
\bibitem[Nardini et al.(2015)]{2015Sci...347..860N} Nardini, E., Reeves, J.~N., Gofford, J., et al.\ 2015, Science, 347, 860 
\bibitem[Veilleux et al.(2013)]{2013ApJ...776...27V} Veilleux, S., Mel{\'e}ndez, M., Sturm, E., et al.\ 2013, ApJ, 776, 27 
\bibitem[Blandford \& Payne(1982)]{1982MNRAS.199..883B} Blandford, R.~D., \& Payne, D.~G.\ 1982, MNRAS, 199, 883 
\bibitem[Proga et al.(2000)]{2000ApJ...543..686P} Proga, D., Stone, J.~M., \& Kallman, T.~R.\ 2000, ApJ, 543, 686 
\bibitem[Fukumura et al.(2010)]{2010ApJ...715..636F} Fukumura, K., Kazanas, D., Contopoulos, I., \& Behar, E.\ 2010, ApJ, 715, 636 
\bibitem[Chartas et al.(2002)]{2002ApJ...579..169C} Chartas, G., Brandt, W.~N., Gallagher, S.~C., \& Garmire, G.~P.\ 2002, ApJ, 579, 169 
\bibitem[Lanzuisi et al.(2012)]{2012A&A...544A...2L} Lanzuisi, G., Giustini, M., Cappi, M., et al.\ 2012, ApJ, 544, A2 
\bibitem[King(2010)]{2010MNRAS.402.1516K} King, A.~R.\ 2010, MNRAS, 402, 1516 
\bibitem[Hopkins \& Elvis(2010)]{2010MNRAS.401....7H} Hopkins, P.~F., \& Elvis, M.\ 2010, MNRAS, 401, 7 
\bibitem[Gaspari et al.(2012)]{2012MNRAS.424..190G} Gaspari, M., Brighenti, F., \& Temi, P.\ 2012, MNRAS, 424, 190 
\bibitem[Wagner et al.(2013)]{2013ApJ...763L..18W} Wagner, A.~Y., Umemura, M., \& Bicknell, G.~V.\ 2013, ApJ, 763, L18 
\bibitem[Ostriker et al.(2010)]{2010ApJ...722..642O} Ostriker, J.~P., Choi, E., Ciotti, L., Novak, G.~S., \& Proga, D.\ 2010, ApJ, 722, 642 
\bibitem[Zubovas \& King(2012)]{2012ApJ...745L..34Z} Zubovas, K., \& King, A.\ 2012, ApJ, 745, L34 
\bibitem[Faucher-Gigu{\`e}re \& Quataert(2012)]{2012MNRAS.425..605F} Faucher-Gigu{\`e}re, C.-A., \& Quataert, E.\ 2012, MNRAS, 425, 605 
\bibitem[Zubovas \& King(2014)]{2014MNRAS.439..400Z} Zubovas, K., \& King, A.~R.\ 2014, MNRAS, 439, 400 
\bibitem[Maiolino et al.(2012)]{2012MNRAS.425L..66M} Maiolino, R., Gallerani, S., Neri, R., et al.\ 2012, MNRAS, 425, L66 
\bibitem[Veilleux et al.(2017)]{2017ApJ...843...18V} Veilleux, S., Bolatto, A., Tombesi, F., et al.\ 2017, ApJ, 843, 18 
\bibitem[Wang et al.(2012)]{2012ApJ...756..180W} Wang, J., Fabbiano, G., Karovska, M., Elvis, M., \& Risaliti, G.\ 2012, ApJ, 756, 180 
\bibitem[Wang et al.(2011)]{2011ApJ...736...62W} Wang, J., Fabbiano, G., Elvis, M., et al.\ 2011, ApJ, 736, 62 
\bibitem[Fischer et al.(2013)]{2013ApJS..209....1F} Fischer, T.~C., Crenshaw, D.~M., Kraemer, S.~B., \& Schmitt, H.~R.\ 2013, ApJ, 209, 1 
\bibitem[Dadina et al.(2010)]{2010A&A...516A...9D} Dadina, M., Guainazzi, M., Cappi, M., et al.\ 2010, A\&A, 516, A9 
\bibitem[Cappi et al.(2013)]{2013arXiv1306.2330C} Cappi, M., Done, C., Behar, E., et al.\ 2013, arXiv:1306.2330 
\bibitem[Waters et al.(2017)]{2017MNRAS.467.3160W} Waters, T., Proga, D., Dannen, R., \& Kallman, T.~R.\ 2017, MNRAS, 467, 3160 
\bibitem[Kaastra et al.(2014)]{2014arXiv1412.1171K} Kaastra, J.~S., Terashima, Y., Kallman, T., et al.\ 2014, arXiv:1412.1171 
\bibitem[Paggi et al.(2017)]{2017ApJ...844....5P} Paggi, A., Kim, D.-W., Anderson, C., et al.\ 2017, ApJ, 844, 5 
\bibitem[Feruglio et al.(2015)]{2015A&A...583A..99F} Feruglio, C., Fiore, F., Carniani, S., et al.\ 2015, A\&A, 583, A99 
\bibitem[Feruglio et al.(2017)]{2017A&A...608A..30F} Feruglio, C., Ferrara, A., Bischetti, M., et al.\ 2017, A\&A, 608, A30 
\bibitem[King \& Pounds(2015)]{2015ARA&A..53..115K} King, A., \& Pounds, K.\ 2015, ARA\&A, 53, 115 
\bibitem[Mizumoto et al.(2018)]{2018arXiv181204316M} Mizumoto, M., Izumi, T., \& Kohno, K.\ 2018, arXiv:1812.04316 
\bibitem[Reeves et al.(2018)]{2018ApJ...854L...8R} Reeves, J.~N., Braito, V., Nardini, E., et al.\ 2018, ApJ, 854, L8 
\bibitem[Pounds \& Vaughan(2011)]{2011MNRAS.413.1251P} Pounds, K.~A., \& Vaughan, S.\ 2011, MNRAS, 413, 1251 
\bibitem[Kormendy \& Kennicutt(2004)]{2004ARA&A..42..603K} Kormendy, J., \& Kennicutt, R.~C., Jr.\ 2004, ARA\&A, 42, 603 
\bibitem[Kormendy(2016)]{2016ASSL..418..431K} Kormendy, J.\ 2016, Galactic Bulges, 418, 431 
\bibitem[{\"O}zel(2018)]{2018NatAs...2..608O} {\"O}zel, F.\ 2018, Nature Astronomy, 2, 608 
\bibitem[Mushotzky(2018)]{2018SPIE10699E..29M} Mushotzky, R.\ 2018, Space Telescopes and Instrumentation 2018: Ultraviolet to Gamma Ray, 10699, 1069929 
\bibitem[Barcons et al.(2017)]{2017AN....338..153B} Barcons, X., Barret, D., Decourchelle, A., et al.\ 2017, Astronomische Nachrichten, 338, 153 
\bibitem[Cicone et al.(2018)]{2018NatAs...2..176C} Cicone, C., Brusa, M., Ramos Almeida, C., et al.\ 2018, Nature Astronomy, 2, 176

\end{document}